\documentclass[12pt]{amsart}

\usepackage[margin=1in]{geometry}
\usepackage[lofdepth,lotdepth,caption=false]{subfig}
\usepackage{fancyhdr}
\usepackage{hyperref}
\usepackage{faktor}
\usepackage{amsmath, amssymb, graphicx}
\usepackage{mathtools}
\usepackage{xspace}
\usepackage{braket}
\usepackage{yfonts}
\usepackage{multicol}
\usepackage{multirow}
\usepackage{color}
\usepackage{setspace}
\usepackage{enumitem}
\usepackage{pst-node}
\usepackage{tikz-cd}
\usepackage{tikz}
\usepackage{tikz-network}
\usetikzlibrary{decorations.markings}
\usepackage{slashed}
\usetikzlibrary{calc}
\usepackage[mathscr]{euscript}
\usepackage[numbers]{natbib}
\usepackage[capitalize,nameinlink]{cleveref}

\newcommand{\cd}{\cdot}
\newcommand{\ra}{\rightarrow}
\newcommand{\pr}{\prime}

\newcommand{\C}{\mathbb{C}}

\newcommand{\R}{\mathbb{R}}

\newcommand{\tld}[1]{\widetilde{#1}}
\newcommand{\lbar}[1]{\overline{#1}}
\newcommand{\id}{\mathrm{id}}

\DeclareMathAlphabet{\mathpzc}{OT1}{pzc}{m}{it}

\newcommand{\mcI}{\mathcal{I}}

\newcommand{\mcL}{\mathcal{L}}

\theoremstyle{plain}

\theoremstyle{definition}
\newtheorem{dfn}{Definition}[section]
\theoremstyle{remark}
\newtheorem{rmk}{Remark}[section]

\begin{document}

\title{The Universe from a Single Particle II}
\maketitle

\begin{center}
  \normalsize
  Michael Freedman \footnote{michaelf@microsoft.com}\textsuperscript{,\hyperref[1]{*}},
  Modjtaba Shokrian Zini \footnote{mshokrianzini@pitp.ca} \textsuperscript{,\hyperref[2]{$\dagger$},\hyperref[3]{$\bullet$}}
 \par   \bigskip
\end{center}

\begin{abstract}
We continue to explore, in the context of a toy model,  the hypothesis that the interacting universe we see around us could result from single particle (undergraduate) quantum mechanics via a novel spontaneous symmetry breaking (SSB) acting at the level of probability distributions on Hamiltonians (rather than on states as is familiar from both Ginzburg-Landau superconductivity and the Higgs mechanism). In an earlier paper \cite{freedman2021universe} we saw qubit structure emerge spontaneously on $\C^4$ and $\C^8$, and in this work we see $\C^6$ spontaneously decomposing as $\C^2\otimes \C^3$ and very curiously $\C^5$ (and $\C^7$) splitting off one (one or three) directions and then factoring. This evidence provides additional support for the broad hypothesis: Nature will seek out tensor decompositions where none are present. We consider how this finding may form a basis for the origins of interaction and ask if it can be related to established foundational discussions such as string theory.
\end{abstract}
\tableofcontents
\section{Introduction}\label{introduction}
This paper continues to explore the genesis of interaction via a novel spontaneous symmetry breaking (SSB) hypothesis. The novelty is that the SSB is on the level of probability distributions on Hermitian operators rather than acting on ground states. In this picture, an undifferentiated Gaussian Unitary Ensemble (GUE) breaks to a quite different probability distribution on Hermitian operators which emphasizes interaction between dof (degrees of freedom) corresponding to relatively low spins, e.g.\ qubits. In this story, unlike inflation \cite{linde15}, the total dimension of the Hilbert space is conserved. In spin language, the breaking is from the physics of a single particle of extremely large spin to a system of interacting particles with small spins.

Using the concept GUE, we obtain \emph{one} toy model for this Hamiltonian of the universe, drawn from the Gaussian ensemble: $\frac{1}{Z}e^{-\mathrm{const.}\operatorname{tr}(H^2)}$, where $\langle A, B \rangle \coloneqq -\operatorname{tr}(iA \ast iB)$ is proportional to the ad-invariant Killing form on the Lie algebra $\operatorname{su}(n)$, where $n = \dim(\text{Hilbert space})$. \emph{Other} toy models are obtained by choosing a functional 
$$f: \{\text{metrics } g_{ij} \text{ on the space } \operatorname{Her}_0(n) \text{ of traceless } n \times n \text{ Hermitian matrices}\} \ra \R,$$
which we think of as a pseudo-energy, and replacing $\operatorname{tr}(H^2)$ with $g_{ij} v^i v^j$ in the formula above, for $g_{ij}$ a local minima of $f$. Such models are Boltzmann in nature, with the probability of the metric $g_{ij}$ being proportional to $e^{-\Delta/k_b T}$, where $\Delta=f(g_{ij})$ and $T$ a pseudo-temperature. The choice of a local minima for $f$ breaks the ad-symmetry of the Killing form.

An alternative possibility, which is numerically more difficult and has not been attempted, is to keep the choice of metric in superposition and exploit interference effects to concentrate that choice near the local minima of $f$. Thus one could associate with $f$ and a real expansion parameter $k$ a normalized metric:
\[
	\lbar{g}_{ij} \coloneqq  \int_{\{g_{ij}\}}\ dvol\ e^{-ikf(g_{ij})}g_{ij}  \Big\slash \det \left\lvert \int_{\{g_{ij}\}}\ dvol\ e^{-ikf(g_{ij})}g_{ij} \right\rvert
\]

In this paper and \cite{freedman2021universe} we have chosen the simplest device: First, a local minima of $f$, called $g_{ij}$, usually quite different from the usual $L^2$-metric $-\operatorname{tr}(iA *i B)$, determines a probability distribution on $\operatorname{Her}_0(n)$, $P_\theta(H)=\frac{1}{Z} e^{-\langle H,H \rangle_{g_{ij}}}, \langle H,K\rangle_{g_{ij}}:= g_{ij}H^i * iK^j$, where $i,j$ run over an $L^2-$orthonormal basis of $\operatorname{Her}_0(n)$\footnote{The space of traceless $n \times n$ Hermitian matrices}. This probability distribution $P_\theta(H)$ is then the source of the Hamiltonian $H_0$ of ``our universe''. $H_0$'s structure is conditioned by $g_{ij}$, in particular by the geometry of the principal axes of $g_{ij}$ relative to the $L^2$-metric. We study these axes, which, of course, should be operators, in search of a qubit or tensor structure. Unlike \cite{freedman2021universe}, we have not restricted $n$ to be a power of 2 and present extensive numerical studies for $n = 4,5,6,7$, and 8. In \cite{freedman2021universe} the central definition was \emph{kaq}, that a metric ``knows about qubits,'' and is reviewed below. For $n=6$ we find metrics which instead know about an isomorphism $\C^6 \cong \C^2 \otimes \C^3$, necessitating an expansion of that concept. Even more remarkable, when $n=5$ and 7, primes, we see local minima where the dimension has been ``rounded down'' to a composite, effectively writing $5 = 2 \times 2 + 1$ and $7 = 2 \times 3 + 1$ (for one local minima) and $7 = 2 \times 2 + 3$ (for another). We will discuss these cases in full detail.

Our choice of functionals $f$ is essentially the same as in \cite{freedman2021universe}; we considered $f =$ Ricci scalar curvature, and also $f =$ real or imaginary parts of perturbed Gaussian integrals built from the symmetric 2-tensor $g_{ij}$ and the structure constants (a 3-tensor) $c_{ij}^k$ for $\operatorname{SU}(n)$. These functionals are also reviewed. Scalar curvature (SC), unlike our other functionals, has no perturbation parameter. So detection of local minima is harder. In this study, no new SC local minima were found beyond those in \cite{freedman2021universe}. But the other functionals yielded many.

There is a historical irony which if not addressed might confuse the reader. We are using the GUE as a model for a random single particle Hamiltonian; the real and imaginary parts of $g_{ij}$, $i>j$, as well as $g_{ii}$ are all i.i.d.\ Gaussian variables. Thus all transitions are equally likely. However, the GUE was introduced by Wigner \cite{wigner55} to model a totality of interactions among many particles which were too complicated to deconstruct, except on the basis of symmetry. Indeed, while statistical spectral properties, such as the semi-circle law, of random interacting Hamiltonians may quickly approach those of the GUE, the more delicate entropic measures which we will introduce shortly distinguish the GUE from interacting ensembles. So for us, the GUE represents random non-interacting Hamiltonians whereas other metrics we find as the \textbf{kaq}-local-minima are manifestly interacting.

Most of our numerical data on SSB comes from Hilbert of dimension $4 \leq n \leq 8$, so caution is in order as we extrapolate to the Hilbert space of the early universe which, if finite, might reasonably be taken\footnote{Derived from \cite{bekenstein73}, larger black holes are estimated to have $\approx 2^{90}$ dof.} to have dimension $n \approx 2^{2^{100}}$, or larger. In our initial study \cite{freedman2021universe} of $n=4,8$ we saw a strong pattern of breaking to qubits. This leads us to speculate that if $n$ is not a power of 2 we would see instead a breaking into qunit tensor factors corresponding to the prime factors of $n$. This line of thought leads to the intriguing possibility that the statistics of prime factors of large random integers would have some residual signature in physical law.\footnote{The Golomb-Dickman constant $\approx 0.624$ is the expected fraction, on a log scale, of the largest prime factor, i.e.\ one would expect the largest prime factor of a thousand digit number to have about 624 digits. Might this constant constrain string theory?} Indeed, our ``discovery'' that $6 = 2 \times 3$, discussed below, validates this thought, call it the ``prime factor scenario.'' However, the further discovery that $5 = 2 \times 2 + 1$ (and that $7 = 2 \times 3 + 1 = 2 \times 2 + 3$) suggest a different, equally interesting, scenario. To be concrete, when $n=5$ we find that there is a fixed decomposition of $\C^5 \cong \C^2 \otimes \C^2 \oplus \C^1$ so that each of the 24 principal axes of a locally minimal $g_{ij}$ assumes a form of a tensor product\footnote{To high numerical precision.} on the first $4 \times 4$ block, with apparently random behavior outside that block. That and similar behavior for $n=7$ suggests symmetry breaking can be to ``almost-\textbf{kaq}'' structures in which a few dimension simply go into a rep\^echage which is ``primordial'' in the sense that it does not seem to participate in many body physics, the physics of the observed world. This scenario might also have implications at low energy, perhaps a tragic one, as overlap of one's wave function with such extraordinary states would not seem salubrious.

As we discuss in the body of this paper, all integers $n$ are well-approximated\footnote{We thank Noam Elkies for this observation. A number is $k$-smooth if it contains only the first $k$-primes in its factorization. A plausible guess for approximation of $n$ by a $k$-smooth number $n_0$ would be $n-n_0\le \mathrm{const} .  \log(n)^{-(k-1)}$.} from below by an integer $n_0$ containing only two primes, say two and three, in their prime factorization: $n - n_0 \leq \mathrm{const.} \log(n)$. So in a Hilbert space $\mathcal{H}$ of dimension about $2^{2^{100}}$ we might expect all but a tiny fraction, about $2^{100}$, of these dimensions to be organized into conventional interacting physics with a rep\^echage of about $2^{100}$ dimensions, still of the primordial non-interacting character: $\mathcal{H} = \mathcal{H}_{\text{int}} \oplus \mathcal{H}_{\text{rep}}$. While the wave function $\psi(t)$ of the universe evolves unitarily on the entire Hilbert space, deviations from unitarity on its projection to $\mathcal{H}_{\text{int}}$ will typically be undetectably small. However, at some time $t$, the projection of $\psi(t)$ to $\mathcal{H}_{\text{rep}}$ of $\psi(t)$, or at least portions of interest to us, could become large, abridging the familiar physical laws. We call this the leaky universe scenario.

Although our findings are empirical, in retrospect they are not  entirely a surprise.  It is well observed in many contexts that extremizing natural functionals leads to solutions exhibiting  striking internal structures and symmetries. For example, in sphere-packing in dimensions 8 and 24, it has been shown that the unique extremal packing ( $E_8$ and Leech respectively)  are, to an extent, independent of the particular choice of energy functional \cite{cohn2019universal}. So the fact that the SSB we observe \textit{creates} exquisitely precise tensor structures as it destroys full rotational symmetry, is not without precedent.

\begin{rmk}\label{bscoredfn1}
As in \cite{freedman2021universe}, our inspiration for examining operator-level SSB was to see if the Brown-Susskind ``penalty metrics'' \cite{brown2017quantum,brown2018second} (where norms decrease exponentially with body number) appeared. Our first chance to look for this is at $n=8$. Indeed we find a local minimum metric with a modest $-0.145$ correlation between norm and a measure of body number we call $b$. Further work, perhaps at $n=16$, will be required to determine the significance of this observation. As explained in \cref{simulationhardness}, $n=16$ is far beyond present methods.
\end{rmk}

Our paper is organized as follows:
\begin{itemize}
	\item Section 2: Reviews the notion of \textbf{kaq} and related concepts.
	\item Section 3: Reviews our choice of functionals and the design of our numerical experiments.
	\item Section 4: Summarizes the totality of our numerical results, including those which previously appeared in \cite{freedman2021universe}.
	\item Section 5: Summary and outlook.
\end{itemize}

\section{Review of \textbf{kaq}}
Our fundamental object of interest is the metric $g_{ij}$ normalized so that $\det(g_{ij}) = 1$ on the linear space of traceless Hermitian $n \times n$ matrices $\operatorname{Her}_0(n)$. Multiplying by $i$ identifies $\operatorname{Her}_0(n)$ with $\operatorname{su}(n)$, traceless skew-Hermitian $n \times n$-matrices, the Lie algebra of $\operatorname{SU}(n)$. Thus $g_{ij}$ (actually $-(g_{ij}iA^i,iB^j) \rotatebox[origin=c]{180}{$\coloneqq$} \langle iA, iB\rangle)$ becomes a metric on $\operatorname{su}(n)$ and a left-invariant metric on $\operatorname{SU}(n)$; this, for example, is in play when we refer to the Ricci scalar curvature as a functional on $\operatorname{Her}_0(n)$. In section 3 we review all the functionals $f$ considered on $\operatorname{Her}_0(n)$. $f$ provides us, numerically, with output a metric $g_{ij}$ where $g_{ij}$ is a local minima for $f$. The question we ask about $g_{ij}$ is whether or not it is ``adapted'' to some decomposition of $\C^n$ into a tensor product of qubits or more generally qunits. We call such adapted metrics \emph{kaq} for ``knows about qubits'' or more generally ``knows about qunits.'' Below we do some dimension counting (and make additional arguments) to show that \textbf{kaq} metrics constitute a subvariety of roughly the square root of the ambient dimension. Startlingly, \textbf{kaq} metrics show up quite regularly and with high numerical precision at local minima for a variety of functionals $f$. Here is the definition.

First, it is easily proven by induction that if $n = p_1 \cd \cdots \cd p_l$ is a prime factorization then: $\operatorname{Her}(n) = \operatorname{Her}(p_1) \otimes \operatorname{Her}(p_2) \otimes \cdots \otimes \operatorname{Her}(p_l)$. Note we have temporarily dropped the traceless condition, and will use the natural inclusion $\operatorname{Her}_0(n) \subseteq \operatorname{Her}(n)$ below to rectify this.

\begin{dfn}
	A qunit structure on $\C^n$ is an equivalence class of $\ast$-isomorphisms $J: \C^{p_1} \otimes \cdots \otimes \C^{p_n} \xrightarrow{\cong} \C^n$ where two are equivalent if related by the left action on the factors by $\operatorname{U}(p_1) \times \cdots \times \operatorname{U}(p_l)$. Thus qunit structures are parameterized by $\operatorname{U}(p_1) \times \cdots \times \operatorname{U}(p_l) \backslash \operatorname{U}(n)$. Note that $J$ induces an isomorphism $j: \operatorname{Her}(p_1) \otimes \cdots \otimes \operatorname{Her}(p_l) \xrightarrow{\cong} \operatorname{Her}(n)$.
\end{dfn}

\begin{dfn}[$\text{\cite[Definition 1.1]{freedman2021universe}}$]\label{kaqdfn}
	A metric $g_{ij}$ on $\operatorname{Her}(n)$ is \textbf{kaq} iff it is \emph{not} ad-invariant, yet there is an isomorphism $j$ (induced from $J$ above) so that $g_{ij}$ possesses a complete set of $n^2-1$ principal axes $\{H_k\}_{1 \leq k \leq n^2-1}$ with
	\[
		H_k = j(H_{1,k} \otimes \cdots \otimes H_{l,k}) \text{ where } H_{s,k} \in \operatorname{Her}(p_s),\ 1 \leq s \leq l
	\]

	In other words, $\C^n$ admits a tensor structure so that the principal axes of $g_{ij}$ ($=$ eigenvectors of $g_i^j$ where the Killing form is used to raise the index) all have compatible tensor structures. Note that $H_k \in \operatorname{Her}_0(n)$, but $H_{s,k} \in \operatorname{Her}(p_s)$.
\end{dfn}

To establish how rare \textbf{kaq} metrics are, consider the following rough dimension count when $n = 2^N$, $N$ large. To specify $J$, and hence $j$, $4^N-1$ parameters are required. To specify each $H_{s,k}$ requires 4 parameters for a total of $4N$, but since scalars pass through the tensor factors $4N$ becomes $3N+1$ for each value of $k$ of which there are $4^N-1$. This makes a total of $(4^N-1) + (3N+1)(4^N-1) = (3N+2)(4^N-1)$ parameters to determine a \textbf{kaq} metric (not normalizing so that $\det(g_{ij}) = 1$), whereas the space of metrics $g_{ij}$ on $\operatorname{su}(2^N)$ has dimension $\frac{4^N(4^N-1)}{2}$ (again without the determinant normalized). Up to log factors, the \textbf{kaq} metrics are asymptotically of square root dimension. Of course since our numerics is for small $N$ we should also investigate $N=2$, where we find equality between the two counts: $8 \cd 15 = 8 \cd 15 = 120$. Clearly we over counted the degrees of freedom in \textbf{kaq}-metrics by treating the principal axes independently. To show that even when $N = 2$ \textbf{kaq} is a proper subvariety we estimate its local dimension around a generic, normalized metric $g_{ij}$ which is diagonal in the so-called \textbf{Pauli-word basis} $\textbf{PB}_n$ defined below.
\begin{dfn}[$\text{\cite[page 3]{freedman2021universe}}$]\label{pauliwordbasis}
We use the Hermitian Pauli operators $I=\begin{pmatrix} 1 & 0 \\ 0 &1 \end{pmatrix}, X=\begin{pmatrix} 0 & 1 \\ 1 & 0 \end{pmatrix}, Y=\begin{pmatrix} 0 & -i \\ i & 0 \end{pmatrix}, Z= \begin{pmatrix} 1 & 0 \\ 0 & -1 \end{pmatrix}$. A \textit{Pauli word} $w$ is an $n-$term tensor product of Pauli operators, such as $I\otimes X\otimes Z\otimes I \otimes I$ for $n=5$. All Pauli words for any $n$ give the Pauli-word basis called $\textbf{PB}_n$. The \textit{weight} of $w$ is the number of non-$I$ letters, which is two in the given example. 
\end{dfn}

We extend $g_{ij}$ from $\operatorname{su}(4)$ to $\operatorname{u}(4)$ by setting $\langle 1 \otimes 1, 1 \otimes 1 \rangle = 1 = g_{0,0}$. So,
\begin{enumerate}
	\item $\det(g) = 1$
	\item $g_{0,0} = 1$
	\item $g_{i,j} = c_i \delta_{i,j}$ where all $c_i$, $1 \leq i \leq 15$, are distinct, and
	\item $\prod_{i=0}^{15} c_i = 1$
\end{enumerate}

Thus $g_{i,j}$, $1 \leq i,j \leq 15$, is our starting point metric on $\operatorname{su}(4)$, and as we vary $g_{i,j}$, the principal axis in the $g_{0,0}$ direction is assumed to stay fixed, of unit norm, and, of course, remains orthogonal to the others.

Consider the three Pauli-word basis elements: $1 \otimes X$, $1 \otimes Y$, and $1 \otimes Z$. As we deform the first, 1 must stay 1, e.g.\ if $1 \otimes X \ra (1 + \delta Y) \otimes (X + \dots)$ no matter how $Y \otimes X$ is deforming perpendicularity will be lost, i.e.\ $\operatorname{tr}((1 + \delta Y)\otimes(X + \dots) \cd (Y + \dots)\otimes (X + \dots)) \neq 0$. So only the left letters $X,Y,Z$ can mix among themselves. Again, $X \ra (aX + bY + cZ)$ is okay but for $d \neq 0$, $X \ra (aX + bY + cZ + d1)$ will break perpendicularity with $1 \otimes 1$. So we see, so far a 3D, $\operatorname{so}(3)$ deformation. Similarly $X \otimes 1$, $Y \otimes 1$, and $Z \otimes 1$ can mix among themselves creating another $\operatorname{so}(3)$ formation. Now consider the 9 weight-2 Pauli words: $X \otimes X, X \otimes Y, \dots, Z \otimes Z$. To maintain a tensor product form and avoid mixing with 1, which would spoil orthogonality with the previous 6 vectors, we see an additional $\operatorname{so}(3) \times \operatorname{so}(3)$ parameter space. So far we have $12 = 3+3+3+3$ parameters.

Now the choice of $J$ (and therefore $j$) is 15 parameters, but the 12 and 15 are \emph{not} independent parameters. They overlap in $\operatorname{so}(3) \times \operatorname{so}(3)$ corresponding to infinitesimal rotations mixing $X$, $Y$, and $Z$ in both left and right factors. The upshot is the local dimensionality near $g = 12 + 15 - 6 = 21 < 120$, showing the \textbf{kaq} variety is proper even when $N=2$.

We suspect that 21 is actually the maximum \textbf{kaq} strata dimension in $\operatorname{Her}_0(4)$.

\section{Loss Functions}
First, we define the loss functions which local minima give us the metrics, and then the loss functions that check their \textbf{kaq}ness.
\subsection{Loss functions to find metrics}~
\\
\indent We review the perturbed Gaussian integral (inspired from \cite{bar1995perturbative}) used to define the functionals in \cite{freedman2021universe}. Let
\begin{align}\label{theintegral}
    & F_k := \int_{\vec{x} \in \R^{3(4^n-1)}}\ \operatorname{d}\vec{x}\ e^{ik(G_{IJ}x^I x^J + c_{ijk}y_1^i y_2^j y_3^k)}
\end{align}
where $x = (y_1,y_2,y_3)$ with $y_o \in \R^{4^n-1}, o \in \{1,2,3\}$, and for $I = (i,o)$, $x^I = y_{o}^i \in \mathbb{R}$, and $G_{IJ}x^I x^J = g_{ij}y_1^i y_1^j + g_{ij} y_2^i y_2^j + g_{ij} y_3^i y_3^j$, i.e. $G = \begin{pmatrix}
    g & 0 & 0 \\
    0 & g & 0 \\
    0 & 0 & g
\end{pmatrix}$. We recall the definition of the structure constants $c_{ij}^k$ of the Lie algebra
\begin{equation}\label{c_ijkdfn}
    [y_i, y_j] = c_{ij}^k y_k \text{ and } c_{ijk} = c_{ij}^{k^\pr} g_{k^\pr k}.
\end{equation}
The real and imaginary part of $F_k$ will be of interest:
\begin{align}
    f_{k,1} = \operatorname{Re}(F_k),\ f_{k,2} = \operatorname{Im}(F_k).
\end{align}
\cref{theintegral} is the most natural \textit{nontrivial} perturbed Gaussian integral from the tensors $g$ and $c$. If instead in \cref{theintegral} we wrote the more obvious integration over $\mathbb{R}^{4^n-1}$, replacing $G$ with $g$ and $x$ with $y$ (\cite[discussion around Eq. (9)]{freedman2021universe}), the skew-symmetry $c_{[ij]k}=0$ would kill the cubic term leaving the Gaussian integral unperturbed. Another functional is derived from the Euclidean version of the above, where $-i$ in the exponent is replaced by $-1$. 

The expansion of the perturbative series of \cref{theintegral} yields a series with the $m$-th term
\begin{align}
    \propto [(c_{ijk} \frac{-i\partial}{\partial V_1^i}\frac{-i\partial}{\partial V_2^j}\frac{-i\partial}{\partial V_3^k})^m e^{-\frac{i}{2}G^{IJ}V_IV_J }]_{\vec{V}=0}.
\end{align}
This can be computed up to third order as in \cite{freedman2021universe} (see also \cite[Equation 1.7 onwards]{bar1995perturbative}). This gives a summation of $m = 2,4,$ and $6$ vertex trivalent tensor networks, where vertices are labelled by $c$ and edges by $g$ or $g^{-1}$.

It is not hard to see that $\operatorname{Im}(F_k) = f_{k,2}$ corresponds to $m \equiv 2 \pmod 4$, while $m \equiv 0 \pmod 4$ gives $\operatorname{Re}(F_k) = f_{k,1}$.  The Euclidean version has alternating signs $\pm1$ depending on $m \equiv 0,2 \pmod  4$. So the numerical experiments are based on $m=2,6$ and on $m = 2,4$ for the Euclidean version. 

Computing the above involves a contraction of a trivalent network without any loops, and $m$ vertices $c_{ijk}$ and edges $g^{ii'},g^{jj'},g^{kk'}$. Furthermore, \cref{c_ijkdfn} implies that vertices can be labelled by $c_{ij}^k$ instead of $c_{ijk}$, while edges are labelled by $g^{ii'},g^{jj'}$ and $g_{kk'}$ instead of $g^{kk'}$.

We refer to \cite[Figures 4-6]{freedman2021universe} for the full list of the tensor diagrams. In \cref{Thetatincanprismdiagram}, we borrow an example from the reference for each $m=2,4,6$:

\begin{figure}[h]
    \centering
\begin{tikzpicture}
\begin{scope}[shift = {(-7,0)}]
\Vertex[x=1,label=$c$]{A}
\Vertex[x=1,y=-2,label=$c$]{B}
\Edge[label=$k$, color = red](A)(B)
\Edge[bend=65,label=$i$](A)(B)
\Edge[bend=-65,label =$j$](A)(B)
\end{scope}
\begin{scope}[shift={(-2,-1.75)}]
\Vertex[label=$c$]{A}
\Vertex[x=2,label=$c$]{B}
\Vertex[y=2,label=$c$]{C}
\Vertex[x=2,y=2,label=$c$]{D}
\Edge[label=$k$, color =red](A)(C)
\Edge[label=$k$, color =red](B)(D)
\Edge[bend=45,label=$i$](A)(B)
\Edge[bend=-45,label=$j$](A)(B)
\Edge[bend=45,label=$i$](C)(D)
\Edge[bend=-45,label=$j$](C)(D)
\end{scope}
\begin{scope}[shift={(3,-1.75)}]
\Vertex[label=$c$]{A}
\Vertex[x=1,y=0.8,label=$c$]{B}
\Vertex[x=0.5,y=2,label=$c$]{C}
\Vertex[x=2,label=$c$]{D}
\Vertex[x=3,y=0.8,label=$c$]{E}
\Vertex[x=2.5,y=2,label=$c$]{F}
\Edge[label=$k$, color =red](A)(B)
\Edge[label=$j$](B)(C)
\Edge[label=$i$](C)(A)
\Edge[label=$k$, color =red](D)(E)
\Edge[label=$j$](E)(F)
\Edge[label=$i$](F)(D)
\Edge[label=$j$](A)(D)
\Edge[label=$i$](B)(E)
\Edge[label=$k$, color =red](C)(F)
\end{scope}
\end{tikzpicture}
    \caption{Theta, tincan and prism diagrams. All diagrams are trivalent networks without any loop, and vertices are the structure constants $c_{ij}^k$. Each vertex has indices $i,j,k$ which are paired with their counterpart in another vertex. This pairing is done using $g$ along edge of type $k$ (colored red) and $g^{-1}$ for type $i$ and $j$. Tincan and prism are given with some sample labeling. Red lines are labelled by $g$ and black lines by $g^{-1}$.}
    \label{Thetatincanprismdiagram}
\end{figure}
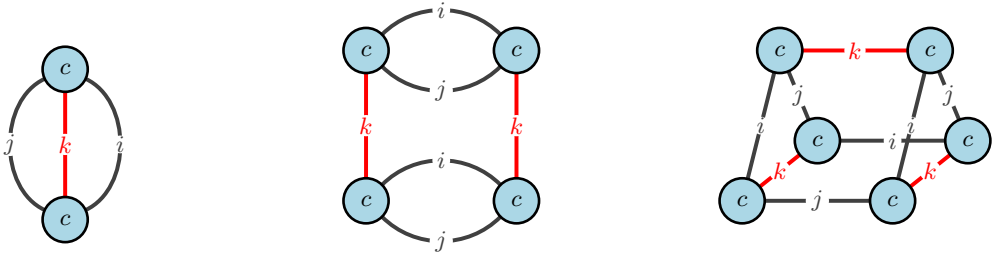

Up to third order, we obtain the following functionals:
\begin{align}\label{F_26}
    F_{26}(c,g,k) = \frac{1}{6!} (\text{sum of 6-vertex diagrams}) - \frac{k^2}{2!}(\text{the 2-vertex diagram Theta})
    \\\label{F_24}
    F_{24}(c,g,k) = \frac{1}{4!}(\text{sum of 4-vertex diagrams}) - \frac{k}{2!} (\text{the 2-vertex diagram Theta}).
\end{align}

\begin{rmk}
Numerical and theoretical evidence shows that each diagram is convex or concave with critical point at $g=Id$ (see \cite[Appendix C]{freedman2021universe}). Hence a signed sum of these diagrams gives local minima around a local maximum at $g=Id$. 
\end{rmk}
To fix the volume with $\det(g) = 1$, we found it to be numerically more stable to take a Lagrangian approach instead of normalizing by $\det(g) = 1$ \cite{freedman2021universe}. Hence we added the term $(\det(g)-1)^2$ with a high enough coefficient to the loss function, giving us the final formulae:
\begin{align}\label{L_24}
    L_{24}(c,g,k) = r_1^{-1}F_{24}(c,g,k) + r_2(\det(g)-1)^2, \\\label{L_26}
    L_{26}(c,g,k) = r_1^{-1}F_{26}(c,g,k) + r_2(\det(g)-1)^2,
\end{align}
where $r_1 \ge 1, r_2 >>1$. The solutions found through gradient descent were experimentally checked to be local minima and these generally have highly degenerate eigenspace. 
\begin{dfn}[$\text{\cite[Definition 2.1]{freedman2021universe}}$]\label{degpatdfn}
The \textit{degeneracy pattern} $(d_1,\ldots,d_t)$ is a tuple describing the dimensions of the eigenspaces ordered by increasing eigenvalues.     
\end{dfn}

\subsubsection{Gradient descent details}\label{Gradient_descent_details2624}
We use Adam gradient descent algorithm in PyTorch with default hyperparameters (learning rate $ = 10^{-3}$), similar to \cite[Section 2.3]{freedman2021universe}. Similarly, we replace $g$ with $g^{-1}$ so that fewer terms involve the calculation of $g^{-1}$, however in the results tables in the next sections, we consider the original eigenvalues of $g$; See \cite[Sections 2.3]{freedman2021universe} for more on the Adam algorithm.

There are many ways to initialize the gradient descent \cite[Section 2.2]{freedman2021universe}. Here, we choose the method \textbf{GenPerturbId} for initialization of $g$: The gradient descent starts at a random metric given by a Gaussian perturbation of the identity metric. If we were to restrict ourselves to only diagonal metrics, \textbf{kaq}ness would have been trivial as the basis is the Pauli word basis (\cite[Theorem 2.1]{freedman2021universe}). 

\subsubsection{Basis}\label{basis}
For $n$ a power of $2$, we use the Pauli-word basis (as in \cite{freedman2021universe}), and for $n=5,6,7$, we use the generalized Gell-Mann basis \cite{gengellman} which is well-known trace orthonormal basis of $\mathfrak{su}(n)$. 

For $n=4,8$, the Pauli-word basis was chosen because of the favorable properties of its structure constants and the implications on the gradient descent (studied in \cite[Appendix A \& B]{freedman2021universe}). For example, the gradient flow starting at a diagonal metric $g$ (\textbf{DiagPerturbId} \cite{freedman2021universe}) will remain in the space of diagonal metrics, thus always giving \textbf{kaq} local minima. As a result, this basis has been shown to provide a better source of \textbf{kaq} examples (see also \cref{genremarksresults}). In this work, we investigate if the gradient flow starting at a nondiagonal metric $g$ (\textbf{GenPerturbId} instead of \textbf{DiagPerturbId}) still delivers \textbf{kaq} examples.

We do not have a Pauli-word basis for $n=5,6,7$, however the well-known Gell-Mann basis coincides with the Pauli-word basis for $\mathfrak{su}(2)$ and the Gell-Mann matrices for $\mathfrak{su}(3)$ (acting on qutrits as the basis for the Gell-Mann's quark model), and can be seen as the generalization of both but acting on qu\textbf{n}its.
\begin{rmk}\label{elemtensorbasis}
In the particular case of $n=6$, we also consider another basis, that can potentially lead to more \textbf{kaq} solutions: $\{$(Gell-Mann basis of $\mathfrak{u}(2)$) $\otimes$ (Gell-Mann basis of $\mathfrak{u}(3)$)$\}-\{Id \otimes Id\}$. We call this basis the \textbf{tensor} basis for $\mathfrak{su}(6)$.
\end{rmk}
\begin{rmk}
We emphasize that, for numerical reasons, the search for (almost-)\textbf{kaq} metrics \textit{is} basis independent, but it is important to search using different well-known bases, especially those which already have a tensor factorization, such as the Pauli-word basis or the tensor basis for $\mathfrak{su}(6)$. As noted before, some of these bases may put the search in a better position to find (almost-)\textbf{kaq} examples.
\end{rmk}

\subsubsection{Why simulate only up to $3$ qubits or $n \le 8$?}\label{simulationhardness}
Current computers are usually capable of simulating quantum computations involving many more than just $3$ qubits. However, we need to evaluate large, $6$-vertex trivalent diagrams. Contraction of these diagrams has to be done in a way to minimize the memory footprint; e.g. the last step of the contraction is always a diagram like Theta, but potentially with more edges. Each edge represents $n^2-1$ indices. Some contractions would end with two vertices with $5$ edges left. This makes the total dimension of these $(n^2-1)^5$. Assuming $4$ qubits, this means $n=16$ and so $255^5 = 1,078,203,909,375$ many parameters. 

For our computations we have to use GPU acceleration as otherwise the runtime of convergence of the gradient descent would be prohibitively long. $4$ billion parameters take about $1$ GB of VRAM on the GPU meaning at least $\sim270$ GB just for storing the parameters of each vertex.  Unlike deep learning computations which can be parallelized across multiple GPUs, our computations require at least each node to be on a single GPU and a single Tesla V100 has only $32$ GB of VRAM, far smaller than $270$. We should emphasize that unlike many quantum computation settings, where matrices are sparse, we do not have this advantage to help us with storage. Even though at the beginning phase, the structure constants are sparse, once a contraction by a general metric $g$ is done, the sparsity is no longer present.

It should also be noted that the requirement for the gradient descent is even larger: Adam gradient descent involves two other quantities (momentum and acceleration) attached to each parameter, making the total VRAM requirement $270\times 3$ GB for each vertex. This is a very low underestimate and the true requirement may be a multiple of that; For example, experiments have shown that $\mathfrak{su}(n=9)$, with $(n^2-1)^5 \sim 3.2 \times 10^9$, requires more than 16GB of VRAM. 
\begin{rmk}\label{whykgrowslarge}
Another fundamental issue we face beyond $n=8$ is that of numerical stability. As $n$ grows, we must also increase $k$ to make it possible for the computer to detect local minima around identity. Otherwise, the value of the functionals $F_{24},F_{26}$ at identity would be too close to the local minima around it for the computer to be able to distinguish them. On the other hand, increasing $k$ makes the value of the functionals too large for the computer to give consistent results, even when using double-precision floating-point.
\end{rmk}

\subsubsection{Ricci scalar curvature}
For all unimodular Lie algebras, the Ricci scalar curvature  is defined as \cite{jensen1971scalar}:
\begin{align}\label{scalarcurvature}
    R = -\frac{1}{2} \sum_{i,i',j,k}c_{ij}^kc_{i'k}^j g_{ii'} -\frac{1}{4}\sum_{i,j,k}\sum_{i',j',k'} c_{ij}^kc_{i'j'}^{k'}g_{ii'}g_{jj'}g^{kk'}
\end{align}

For $\mathfrak{su}(2^n)$, this simplifies to
\begin{align}\label{Rformula}
    R = \frac{1}{2} \sum_{i,j,k}(c_{ij}^{k})^2 g_{ii} -\frac{1}{4}\sum_{i,j,k}\sum_{i',j',k'} c_{ij}^kc_{i'j'}^{k'}g_{ii'}g_{jj'}g^{kk'}.
\end{align}
See \cite[Figure 7]{freedman2021universe} for the diagrammatic formulation. As mentioned in \cite{freedman2021universe}, to find critical points, $||\nabla R||$ must be minimized. This gradient can be computed explicitly using \cref{Rformula}. Gradient descent on this problem does not work and evolutionary algorithms must be used \cite{freedman2021universe}. However we report that no new local minima were found even after a more comprehensive evolutionary search. 

\subsection{Loss functions for checking (almost-)\textbf{kaq}ness}\label{losskaqnesssection}~
\\
\indent Once solutions in the previous section are found, we would like to check if they are (almost-)\textbf{kaq}. Using entropy, one can define a loss function which evaluates to zero if and only if the solution found is \textbf{kaq}. The question is: what are the parameters of our loss function? 

Let $g$ be a solution for any of the above loss functions, and let the eigenbasis of $g$ be $\{iH_1, \ldots, iH_{n^2-1}\}$ where all are normalized in $l_2$-norm. There are two sources for the parameters. The first set of parameters describe the conjugation of the eigenbasis by some $U\in U(n)$, which is what describes the function $j$ in \cref{kaqdfn}. However the choice of each eigenspace basis is not unique, specifically, every degenerate eigenspace of degree $d$ can afford an independent change of basis. Thus every such eigenspace gives additional parameters describing an orthogonal matrix $V \in O(d)$. So the number of parameters is $n^2 + \sum_{i=1}^{t} (d_i^2-d_i)/2$ where $(d_1,\ldots,d_t)$ is the degeneracy pattern of $g$. We use $\theta$ to denote all these parameters.

\subsubsection{Computing entropies}
After the above two transformations, by abuse of notation, let the new orthonormal eigenbasis be $\{iH_1, \ldots, iH_{n^2-1}\}$. Then viewing each matrix $H_j$ as a $n^2\times1$ vector $ v_{j}$, and given $n=\prod_{i=1}^l p_i$, we compute the entropy $s_{ij}(g,\theta)$ for $1\le i \le l, 1\le j \le n^2-1$.

This is the entropy for the decomposition $v_{j} = v_{j,i,1} \otimes v_{j,i,2}$ where $v_{j,i,1} \in  \C^{p_i} \otimes (\C^{p_i})^*$ and $v_{j,i,2} \in \C^{n/p_i} \otimes (\C^{n/p_i})^*$ for $1 \le i \le l$. We compute this entropy by taking the Schmidt decomposition $v_{j} = \sum_l \alpha_{jil} w_{j,i,l,1}\otimes w_{j,i,l,2}$, where $w_{j,i,l,1} \in  \C^{p_i} \otimes (\C^{p_i})^*$ and $w_{j,i,l,2} \in \C^{n/p_i} \otimes (\C^{n/p_i})^*$, and so
$$s_{ij}(g,\theta) = \sum_l -|\alpha_{jil}|^2\log(|\alpha_{jil}|^2).$$
The Schmidt decomposition provides the bonus of also having the best candidate $v_{j,i,1} \otimes v_{j,i,2}$ for the tensor decomposition of $H_j$, i.e. $v_{j,i,1} = w_{j,i,o,1}, v_{j,i,2}= w_{j,i,o,2}$ for $o = \text{argmax}_l |\alpha_{jil}|$.

\subsubsection{\textbf{kaq} loss function}
Summing up the entropies gives the \textbf{kaq} loss function
\begin{align}
    \mcL_{\textbf{kaq}}(g,\theta) = \sum_{\substack{1\le i \le l \\ 1\le j \le n^2-1}} s_{ij}(g,\theta).
\end{align}

\begin{rmk}\label{partialkaqdfn}
In the case of $n=8=2\times2\times2$, we will see sometimes \textit{partial-\textbf{kaq}} decomposition into $\C^4 \otimes \C^2$ instead of a full qubit decomposition, by taking the loss function as $\sum_{1\le j \le n^2-1} s_{1j}(g,\theta)$ or $\sum_{1\le j \le n^2-1} s_{2j}(g,\theta)$ or $\sum_{1\le j \le n^2-1} s_{3j}(g,\theta)$.
\end{rmk}
\begin{rmk}\label{sjrmk}
With the exception of $n=8$ which has three primes in its prime decomposition, all other examples of (almost-)\textbf{kaq} have only two. Therefore, we sometimes use $s_j$ as $s_{1j}=s_{2j}$.
\end{rmk}
\subsubsection{Almost-\textbf{kaq} loss function}\label{almostkaqrmk}
It is clear how the above loss function can be altered for cases like $n=5$ ($n=7$), when one might be interested in a decomposition of the form $\C^2 \otimes \C^2 \oplus \C$ ($\C^2 \otimes \C^3 \oplus \C$). This can be done by computing the entropy for the $4\times 4$ ($6\times 6$) upper-left block of each $H_j$, while also adding the norm squared of the 8 (12) entries outside the blocks to the loss function to encourage a block diagonal decomposition.

\subsubsection{Gradient descent details}\label{graddesdetailskaqness}
We had to use the less sophisticated SGD (Stochastic Gradient Descent) algorithm with learning rate 1e-3 and momentum 0.9 for the gradient descent (other hyperparameters were set as their default in \href{https://pytorch.org/docs/stable/generated/torch.optim.SGD.html}{PyTorch}). The alternative (Adam) was found to have issues, likely due to the fact that PyTorch has recently been updated to include gradient descent on real-valued functionals with \textit{complex} parameters. 

\begin{rmk}\label{kaqnessthreshold}
We observed that \textbf{kaq}ness had either strong indication of being present with $\max_{i,j} s_{ij} \sim 10^{-3}$, or otherwise, where mostly $\max_{i,j} s_{ij} > 0.5$. Thus three orders of magnitude typically separate positive from our negative finding of \textbf{kaq}ness.
\end{rmk}
\begin{rmk}[Tolerance margin]\label{tolerancemargin}
To distinguish between different eigenvalues, we used a \textit{tolerance} margin of 0.02 (before changing $g^{-1}$ back to $g$). Hence, in our searches, we gathered the eigenvalues that were the same up to 0.02  as corresponding to the same eigenspace. This choice was made by observing that sometimes solutions with very close spectrum had slight differences (order of 1e-2) in their eigenvalues. Note that a higher tolerance margin increases the number of parameters $\theta$ in $\mathcal{L}_\textbf{kaq}(g,\theta)$,  as it increases the degeneracy dimensions, thereby potentially increasing the chances of getting \textbf{kaq} solutions.
\end{rmk}

\section{Solutions of \texorpdfstring{$F_{24},F_{26}$}{} and their (almost-)\textbf{kaq}ness}
\subsection{Values of \texorpdfstring{$k,r_1,r_2$}{}}~
\\
\indent We shall first list the values of $k,r_1,r_2$ chosen for $n=5,6,7$ in our experiments. As mentioned in \cref{Gradient_descent_details2624}, the gradient descent initialization in all cases is \textbf{GenPerturbId}. For $n=4,8$ we refer to \cite[Tables 1-2 (\textbf{GenPerturbId})]{freedman2021universe}. We note that for $n=6$, as mentioned in \cref{elemtensorbasis}, we have two different bases: the usual Gell-Mann basis, and the \textbf{tensor} basis coming from $\mathfrak{u}(2) \otimes \mathfrak{u}(3)$. The second table in each of \cref{scalingfactorstable24} and \cref{scalingfactorstable26} are for the tensor basis.

\begin{table}[h]
    \centering
    \begin{center}
    \begin{tabular}{ |c|c| } 
    \hline
    $\mathfrak{su}(5)$ & $k=200,400$  \\
    \hline
    $r_1$ & $10$ \\ 
    $r_2$ & $10^4$ \\
    \hline
    \end{tabular}
    \begin{tabular}{ |c|c|} 
    \hline
    $\mathfrak{su}(6)$ (\textbf{tensor} basis) & $k=300,600$ \\
    \hline
    $r_1$ & $10^2$ \\ 
    $r_2$ & $10^5$ \\
    \hline
    \end{tabular}
    \begin{tabular}{ |c|c|} 
    \hline
    $\mathfrak{su}(6)$ & $k=300,600$ \\
    \hline
    $r_1$ & $10$ \\ 
    $r_2$ & $10^4$ \\
    \hline
    \end{tabular}
    \begin{tabular}{ |c|c| } 
    \hline
    $\mathfrak{su}(7)$ & $k=400,800$ \\
    \hline
    $r_1$ & $10$ \\ 
    $r_2$ & $10^4$ \\
    \hline
    \end{tabular}
    \end{center}
    \caption{The scaling factors for $L_{24}$.}
    \label{scalingfactorstable24}
\end{table}
\begin{table}[h]
    \centering
    \begin{center}
    \begin{tabular}{ |c|c| } 
    \hline
    $\mathfrak{su}(5)$ & $k=200,400$ \\
    \hline
    $r_1$ & $10^2$ \\ 
    $r_2$ & $10^5$ \\
    \hline
    \end{tabular}
    \begin{tabular}{ |c|c|} 
    \hline
    $\mathfrak{su}(6)$ (\textbf{tensor} basis) & $k=300,600$ \\
    \hline
    $r_1$ & $10^2$ \\ 
    $r_2$ & $10^5$ \\
    \hline
    \end{tabular}
    \begin{tabular}{ |c|c|} 
    \hline
    $\mathfrak{su}(6)$ & $k=300,600$ \\
    \hline
    $r_1$ & $10^2$ \\ 
    $r_2$ & $10^5$ \\
    \hline
    \end{tabular}
    \begin{tabular}{ |c|c| } 
    \hline
    $\mathfrak{su}(7)$ & $k=400,800$  \\
    \hline
    $r_1$ & $10^4$ \\ 
    $r_2$ & $10^4$ \\
    \hline
    \end{tabular}
    \end{center}
    \caption{The scaling factors for $L_{26}$.}
    \label{scalingfactorstable26}
\end{table}

\subsection{Degeneracy patterns and (almost-)\textbf{kaq}ness}~
\\
\indent We list the local minima found in our search by their degeneracy patterns and their (almost-)\textbf{kaq}ness, as we did in \cite{freedman2021universe}. For each of the 8 tables listed above, we ran the simulation for 15 different random seeds. For $n=4,8$, there are also around 15 simulations for each configuration. We shall also list again the patterns found in \cite{freedman2021universe} for $n=4,8$, this time with their \textbf{kaq}ness specified. In doing so, we note that some of the patterns borrowed from \cite[Section 3.1]{freedman2021universe} for $n=4,8$ do not reappear exactly as they were; for example, (1,3,1,8,2) reappears as (1,4,8,2) in \cref{l24results4}. This discrepancy is due to taking a different (higher) tolerance margin for declaring ``degeneracy'' (\cref{tolerancemargin}).

\subsubsection{Remarks on the presentation of the results}\label{rmkpresentation}
\begin{enumerate}
    \item Within the description and captions, we will use $(d_i)$ for the degeneracy pattern (\cref{degpatdfn}) and thus, $d_i$ refers to the dimension of an eigenspace.
    \item In all tables, we mention the number of different patterns. 
    \item In some of the tables, we have to give some explanation on the solutions and their (almost-)\textbf{kaq}ness. 
    \item In some cases a lot of different patterns are found, in which case, we mention the \textit{best} example(s), e.g. ones that are \textbf{kaq} with $\max_{ij} s_{ij} \sim 10^{-3}$ or the closest example to almost-\textbf{kaq} in terms of the value of the loss function. 
    \item In some other cases, like in \cref{l24results4}, only a few patterns are found and we list them as ``$(d_1,\ldots,d_t): x/y$'' meaning $x$ solutions out of the $y$ solutions with pattern $(d_1,\ldots,d_t)$ are (almost-)\textbf{kaq}. Furthermore, as mentioned before in \cref{kaqnessthreshold}, we have $\max_{ij} s_{ij} \sim 10^{-3}$ in such cases.
    \item Some tables (like Tables \ref{l24results7}-\ref{l24results8}) only show solutions for a single (higher) value of $k$. In all such cases, the lower value gave solutions very close to identity (see \cref{whykgrowslarge}), so we decided not to include them.
\end{enumerate}

In the next section, we will draw some conclusions on the results. We list the results below.

\begin{table}[h]
    \centering
    \begin{center}
    \begin{tabular}{ |p{8cm}|p{8cm}| } 
    \hline
    $k=100$ & $k=200$ \\
    \hline
    $(10,5)$: 3/3  \newline $(1,4,8,2)$: 13/14  &  $(1,1,2,4,2,1,2,2)$: 4/7 \newline $(1,3,1,8,2)$: 5/6  \\
    \hline 
    \end{tabular}
    \end{center}
    \caption{\textbf{Kaq}ness for $L_{24}$ on $\mathfrak{su}(4)$.}
    \label{l24results4}
\end{table}
\begin{table}[h]
    \centering
    \begin{center}
    \begin{tabular}{ |p{8cm}|p{8cm}| } 
    \hline
    $k=200$ & $k=400$ \\
    \hline
    15 patterns. Best pattern (3, 2, 6, 2, 2, 2, 5, 1, 1) has $\max_j s_j = 0.344$ and $mean(s_j)=0.06$, indicating a fairly precise tensor structure. The average of the norm squared of entries outside the blocks is $\sim 0.046$. Other patterns with maximum entropy of 0.49 and 0.69 are present as well. & 12 patterns. Best pattern (1, 1, 1, 10, 2, 2, 2, 1, 2, 2) appears twice with $\max_j s_j = 0.064$ and $mean(s_j)=0.011$, also a fairly precise tensor structure. The average of the norm squared of entries outside the blocks is $\sim 0.044$. Other patterns with maximum entropy of 0.075 and 0.69 are present as well. \\
    \hline 
    \end{tabular}
    \end{center}
    \caption{Almost-\textbf{kaq}ness for $L_{24}$ on $\mathfrak{su}(5)$. See \cref{sjrmk} and \cref{almostkaqrmk} for how to compute entropy $s_{j}(g,\theta)$ for $\C^2\otimes \C^2 \oplus \C$. Note that the average of a random entry from an $l_2$-normalized $5\times 5$ matrix is $0.04$ (to be compared with above).}
    \label{l24results5}
\end{table}
\begin{table}[h]
    \centering
    \begin{center}
    \begin{tabular}{ |p{8cm}|p{8cm}| } 
    \hline
    \multicolumn{2}{|c|}{\textbf{kaq} decomposition to $\C^2 \otimes \C^3$ with \textbf{tensor} basis (\cref{elemtensorbasis})} \\
    \hline
    $k=300$ & $k=600$ \\
    \hline
    15 patterns with very small $d_i$s; none were \textbf{kaq} as minimum $\max_j s_j$ among all solutions was 0.9.  &  15 patterns, one without any degeneracy. None were \textbf{kaq} as minimum $\max_j s_j$ among all solutions was $>$ 1. \\
    \hline 
    \hline
    \multicolumn{2}{|c|}{\textbf{kaq} decomposition to $\C^2 \otimes \C^3$ with Gell-Mann basis} \\
    \hline
    $k=300$ & $k=600$ \\
    \hline 
    13 patterns with overall better degeneracy than above, but none were \textbf{kaq} as minimum $\max_j s_j$ was 0.69. &  Four patterns found. None were \textbf{kaq}: \newline (1, 1, 3, 1, 2, 4, 2, 8, 8, 1, 2, 2): 0/2 \newline (1, 1, 3, 1, 2, 4, 2, 8, 4, 4, 1, 2, 2): 0/2 \newline (1, 1, 3, 1, 6, 2, 8, 8, 1, 2, 2): 0/8 \newline (1, 1, 3, 1, 4, 2, 2, 8, 8, 1, 2, 2): 0/3  \\
    \hline 
    \end{tabular}
    \end{center}
    \caption{\textbf{Kaq}ness for $L_{24}$ on $\mathfrak{su}(6)$. We note how it becomes harder to find solutions with higher degeneracy dimensions $d_i$s, hence decreasing the dof of $\mathcal{L}_\textbf{kaq}$ to find \textbf{kaq}ness.}
    \label{l24results6}
\end{table}
\begin{table}[h]
    \centering
    \begin{center}
    \begin{tabular}{ |p{8cm}|p{8cm}| } 
    \hline
    \multicolumn{2}{|c|}{almost-\textbf{kaq} decomposition to $\C^2 \otimes \C^3\oplus \C$} \\
    \hline
    \multicolumn{2}{|c|}{$k=800$} \\
    \hline
    \multicolumn{2}{|p{16cm}|}{9 patterns found. (1, 1, 8, 1, 4, 4, 12, 12, 1, 2, 2) found twice, where in the best case with strong almost-\textbf{kaq} signal has $\max_j s_j = 0.048$ and $mean(s_j)=0.013$. The average of the norm squared of entries outside the blocks is 0.023.}\\
    \hline 
    \hline
    \multicolumn{2}{|c|}{almost-\textbf{kaq} decomposition to $\C^2 \otimes \C^2\oplus \C^3$} \\
    \hline
    \multicolumn{2}{|c|}{$k=800$} \\
    \hline
    \multicolumn{2}{|p{16cm}|}{Solutions set is the same as above (9 patterns). Two patterns with strong almost-\textbf{kaq} signal: \newline (1, 9, 1, 8, 12, 12, 1, 2, 2) has $\max_j s_j = 0.035$ and $mean(s_j)=0.005$ and average norm squared of entries outside the blocks being $0.026$.  \newline (1, 9, 1, 4, 4, 12, 12, 1, 2, 2) found four times, once with $\max_j s_j = 0.032$ and $mean(s_j)=0.003$ and average norm squared of entries outside the blocks also being $0.026$.}\\
    \hline
    \end{tabular}
    \end{center}
    \caption{Almost-\textbf{kaq}ness for $L_{24}$ on $\mathfrak{su}(7)$. Note that the average of a random entry from an $l_2$-normalized $7\times 7$ matrix is $0.02$ (to be compared with above). Note that there are 24 entries outside the $7\times 7$ matrix blocks for $\C^2 \otimes \C^2\oplus \C^3$ decomposition, and 12 for $\C^2 \otimes \C^3\oplus \C$.}
    \label{l24results7}
\end{table}
\begin{table}[h]
    \centering
    \begin{center}
    \begin{tabular}{ |p{8cm}| } 
    \hline
     $k=1000$ \\
    \hline
    $(1,16,1,6,2,16,16,1,2,2)$: 7/13 - 0/13 \\
    \hline 
    \end{tabular}
    \end{center}
    \caption{\textbf{Kaq}ness for $L_{24}$ on $\mathfrak{su}(8)$. Format is ``$(d_i)$ : Fraction of solutions with the pattern $(d_i)$ that were (partially-\textbf{kaq}) - (\textbf{kaq})''. See \cref{partialkaqdfn} for partial-\textbf{kaq}.}
    \label{l24results8}
\end{table}
\begin{table}[h]
    \centering
    \begin{center}
    \begin{tabular}{ |p{8cm}|p{8cm}| } 
    \hline
    $k=100$ & $k=200$ \\
    \hline 
    $(10,5)$: 3/3  \newline $(3,1,1,8,2)$: 9/11 &  $(10,5)$: 1/1  \newline $(8,6,1)$: 0/1  \newline $(3,1,4,2,1,4)$: 0/7  \newline $(1,3,1,4,4,2)$: 0/6\\
    \hline 
    \end{tabular}
    \end{center}
    \caption{\textbf{Kaq}ness for $L_{26}$ on $\mathfrak{su}(4)$.}
    \label{l26results4}
\end{table}
\begin{table}[h]
    \centering
    \begin{center}
    \begin{tabular}{ |p{8cm}|p{8cm}| } 
    \hline
    $k=200$ & $k=400$ \\
    \hline
    Two patterns (10, 1, 8, 5) and (8, 1, 1, 6, 6, 2) found. The latter with $\max_j s_j = 0.693$ and $mean(s_j)=0.2$. The average of the norm squared of entries outside the blocks is $\sim 0.051$. &  5 patterns. (1, 3, 8, 5, 4, 3) appeared 8 times with $\max_j s_j < 1e-3$. The average of the norm squared of entries outside the blocks is $\sim 0.055$. \\
    \hline 
    \end{tabular}
    \end{center}
    \caption{Almost-\textbf{kaq}ness for $L_{26}$ on $\mathfrak{su}(5)$. See \cref{sjrmk} and \cref{almostkaqrmk} for how to compute entropy $s_{j}(g,\theta)$ for $\C^2\otimes \C^2 \oplus \C$. It should be noted that the average of a random entry from an $l_2$-normalized $5\times 5$ matrix is $0.04$ (to be compared with the numbers above).}
    \label{l26results5}
\end{table}
\begin{table}[h]
    \centering
    \begin{center}
    \begin{tabular}{ |p{8cm}|p{8cm}| } 
    \hline
    \multicolumn{2}{|c|}{\textbf{kaq} decomposition to $\C^2 \otimes \C^3$ with \textbf{tensor} basis (\cref{elemtensorbasis})} \\
    \hline
    $k=300$ & $k=600$ \\
    \hline
    (10, 3, 1, 16, 5): 7/15   & 8 patterns found. Two \textbf{kaq}s: \newline
    (1, 3, 3, 12, 9, 4, 3): 1/2 \newline (10, 3, 1, 8, 8, 5): 1/1 \\
    \hline 
    \hline
    \multicolumn{2}{|c|}{\textbf{kaq} decomposition to $\C^2 \otimes \C^3$ with Gell-Mann basis} \\
    \hline
    $k=300$ & $k=600$ \\
    \hline
    (10, 3, 1, 16, 5): 3/15  &  7 patterns found. Only \textbf{kaq} pattern:\newline
    (1, 3, 3, 6, 6, 9, 4, 3): 2/2 \\
    \hline 
    \end{tabular}
    \end{center}
    \caption{\textbf{Kaq}ness for $L_{26}$ on $\mathfrak{su}(6)$. To be compared with \cref{l24results6}.}
    \label{l26results6}
\end{table}
\begin{table}[h]
    \centering
    \begin{center}
    \begin{tabular}{ |p{8cm}|p{8cm}| } 
    \hline
    \multicolumn{2}{|c|}{almost-\textbf{kaq} decomposition to $\C^2 \otimes \C^3\oplus \C$} \\
    \hline
    $k=400$ & $k=800$ \\
    \hline
    7 patterns with (10, 8, 1, 24, 5) appearing 9 times. Best pattern (18, 1, 24, 5) with $\max_j s_j = 0.06$, $mean(s_j)=0.014$ and average norm squared of entries outside the block $0.025$, showing a strong almost-\textbf{kaq} indication. &  15 different patterns with small $d_i$s and no pattern close to being almost-\textbf{kaq}. \\
    \hline 
    \hline
    \multicolumn{2}{|c|}{almost-\textbf{kaq} decomposition to $\C^2 \otimes \C^2\oplus \C^3$} \\
    \hline
    $k=400$ & $k=800$ \\
    \hline
    Same solutions as above (7 patterns). Best pattern (10, 8, 1, 24, 5) with smallest $\max_j s_j = 0.051$, $mean(s_j)=0.011$ and average norm squared of entries outside the blocks also being $0.022$, showing a strong almost-\textbf{kaq} signal. & Same as above. \\
    \hline
    \end{tabular}
    \end{center}
    \caption{Almost-\textbf{kaq}nes for $L_{26}$ on $\mathfrak{su}(7)$. Note that there are 24 entries outside the matrix blocks for the $\C^2 \otimes \C^2\oplus \C^3$ decomposition, and 12 for $\C^2 \otimes \C^3\oplus \C$.}
    \label{l26results7}
\end{table}
\begin{table}[h]
    \centering
    \begin{center}
    \begin{tabular}{ |p{8cm}|p{8cm}| } 
    \hline
    $k=500$ & $k=1000$\\
    \hline
    $(10,15,1,32,5)$ : 17/17 - 11/17 & $(1,9,1,12,16,9,8,4,3)$: 5/24 - 0/24  \newline $(10,15,1,32,5)$:  1/1 - 1/1  \newline $(1,3,8,1,8,18,5,12,4,3)$: 0/1 - 0/1 \newline $(35,1,24,3)$:  1/1 - 0/1\\
    \hline 
    \end{tabular}
    \end{center}
    \caption{\textbf{Kaq}ness for $L_{26}$ on $\mathfrak{su}(8)$. Format is ``$(d_i)$ : Fraction of solutions with the pattern $(d_i)$ that were (partially-\textbf{kaq}) - (\textbf{kaq})''. The last two patterns in $k=1000$ were found upon further search and are not in \cite{freedman2021universe}.}
    \label{l26results8}
\end{table}

\clearpage

\subsection{Remarks on the results}\label{genremarksresults}~
\\
\indent We make a few general remarks regarding the results.
\begin{enumerate}
    \item The diversity of the degeneracy pattern is lesser for $L_{26}$ compared to $L_{24}$.
    \item (Almost-)\textbf{Kaq}ness is more present in $L_{26}$ solutions than in $L_{24}$'s.
    \item As $k$ increases, the $d_i$s become smaller in dimension, which should make (almost-)\textbf{kaq}ness generally harder to find, as there are less parameters for $\mathcal{L}_{\textbf{kaq}}$ to play with to reach the value of 0 (see \cref{losskaqnesssection} for parameters count). This issue can be seen e.g. in \cref{l26results7} for $k=800$.
    \item The $d_i$s are also smaller for $n$ not power of two. In some cases most or all $d_i=1$, especially for $L_{24}$. This decreases the dof of $\mathcal{L}_\textbf{kaq}$ to find (almost-)\textbf{kaq}ness.
    \item There are also more degeneracy patterns found for both $L_{24},L_{26}$ when $n$ is not a power of two, so much that we did not list all patterns found in some cases.
    \item With regards to almost-\textbf{kaq}ness, we see that in most cases the entries outside the blocks have norm close to that of a random entry from a hermitian matrix of the same size. Therefore, even though the upper-left block is very close to a tensor form, the entries outside of the blocks have not been completely suppressed to zero.
    \item The basis used for $n=4,8$ has been the Pauli-word basis \cite{freedman2021universe}. Changing this basis to the Gell-Mann basis did not give any new pattern or \textbf{kaq}ness result.
\end{enumerate}

\subsection{Lie subalgebras among (almost/partial-)\textbf{kaq} solutions}
\begin{figure}[h]
    \centering
    \includegraphics[trim=5cm 0 0 0, scale = 0.25]{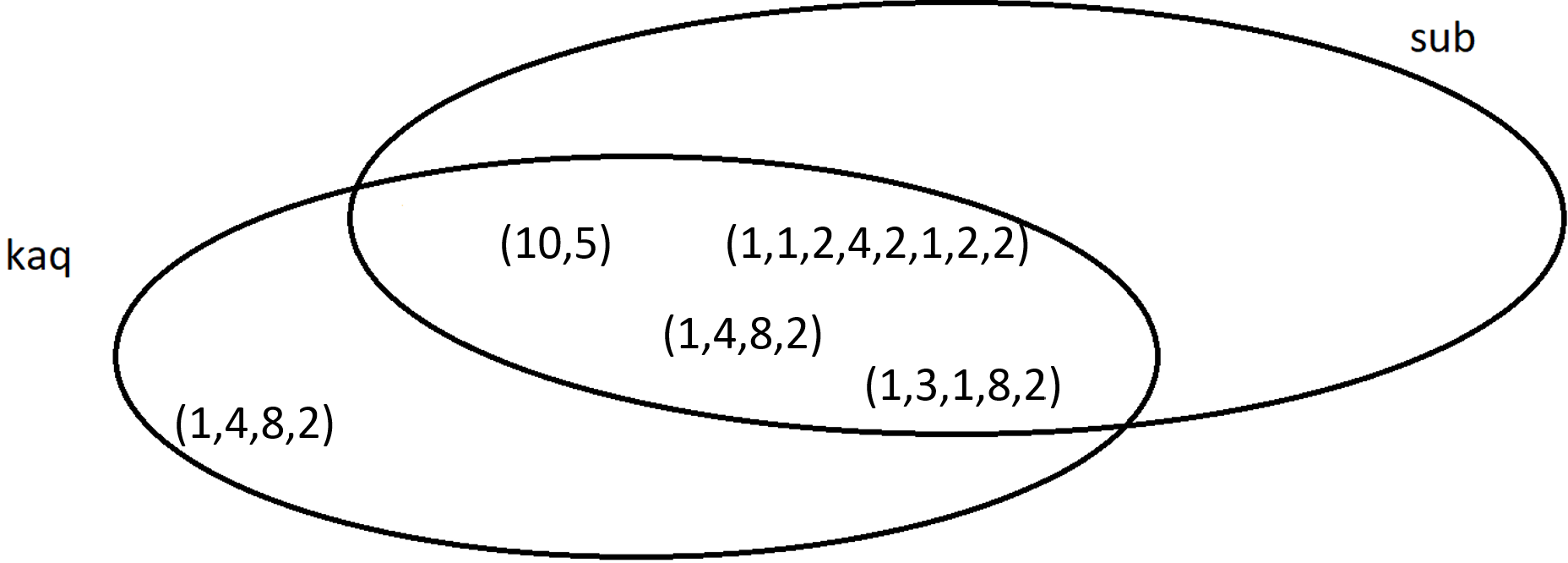}
    \includegraphics[trim=-5cm 0 0 0, scale = 0.25]{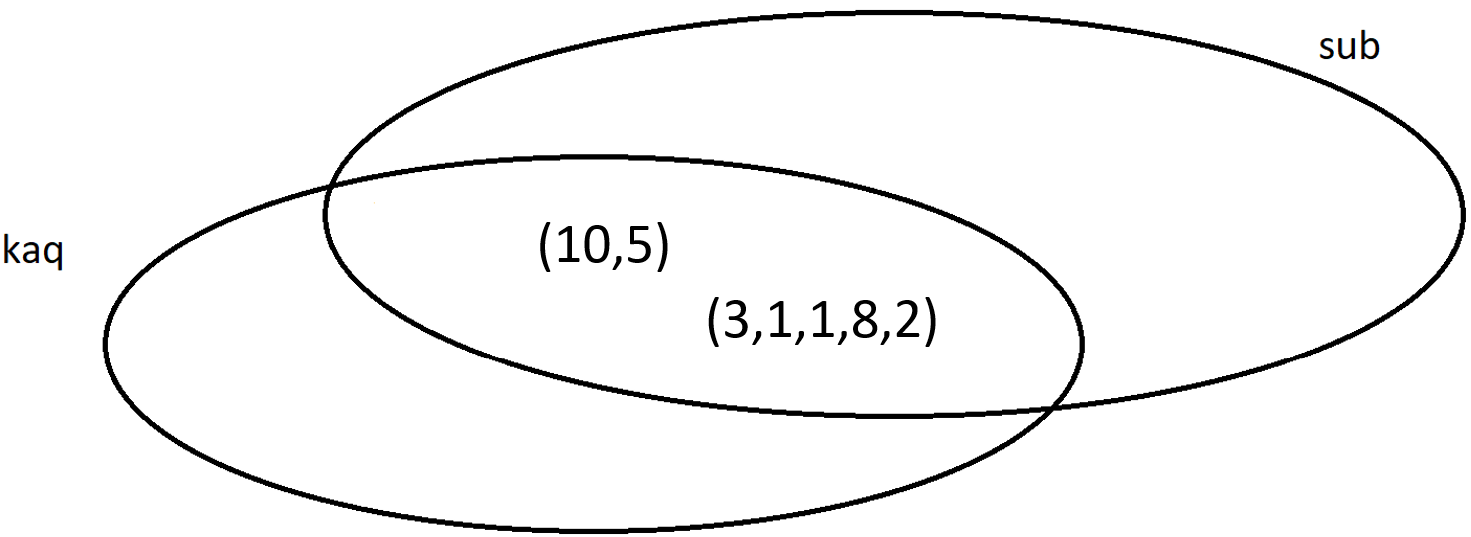}
    \caption{$\mathfrak{su}(4)$: Of the 13 \textbf{kaq} instances with pattern $(1,4,8,2)$, ten of them are not \textbf{sub}. The other three are \textbf{sub} by the combination of the $2D+1D$ eigenspaces forming a $3D$ Lie subalgebra ($\mathfrak{su}(2)$). The same is true for $(1,1,2,4,2,1,2,2)$ where the second $1D$ along with the next $2D$ form an $\mathfrak{su}(2)$. The $10D$ eigenspace in $(10,5)$ also gives a Lie subalgebra (likely isomorphic to $\mathfrak{sp}(2)$ as observed in \cite{freedman2021universe}).}
    \label{l246su4}
\end{figure}

\begin{figure}[h]
    \centering
    \includegraphics[trim=5cm 0 0 0, scale = 0.25]{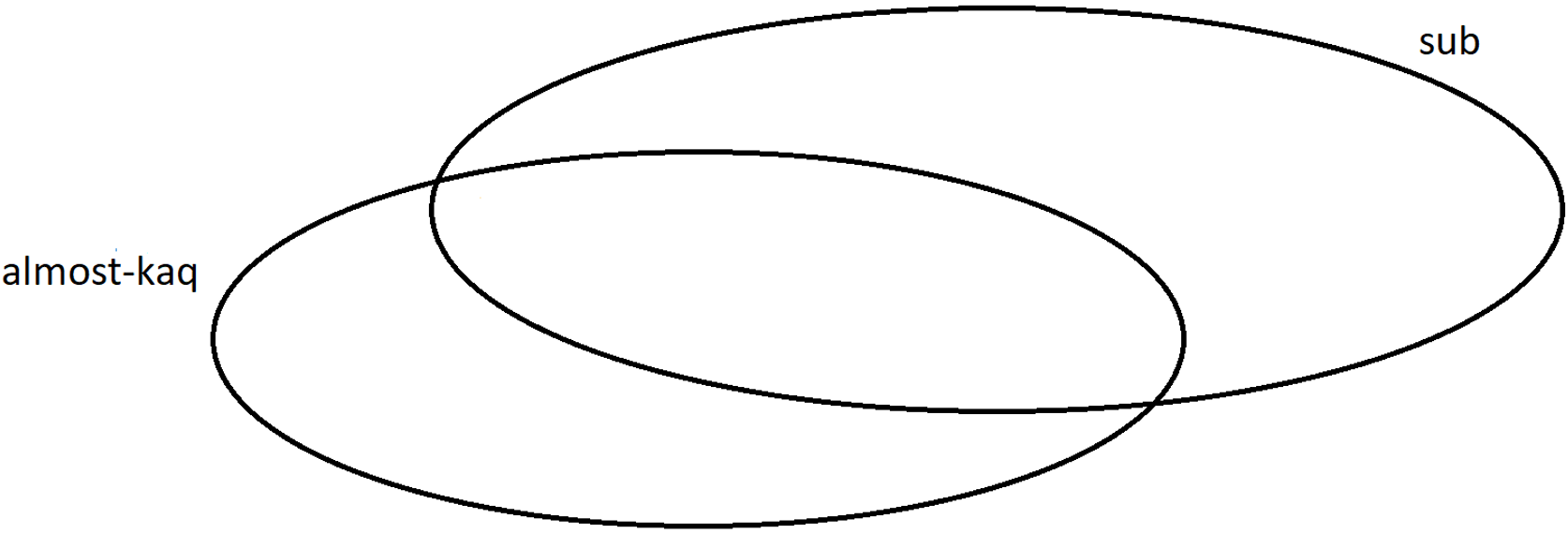}
    \includegraphics[trim=-5cm 0 0 0, scale = 0.25]{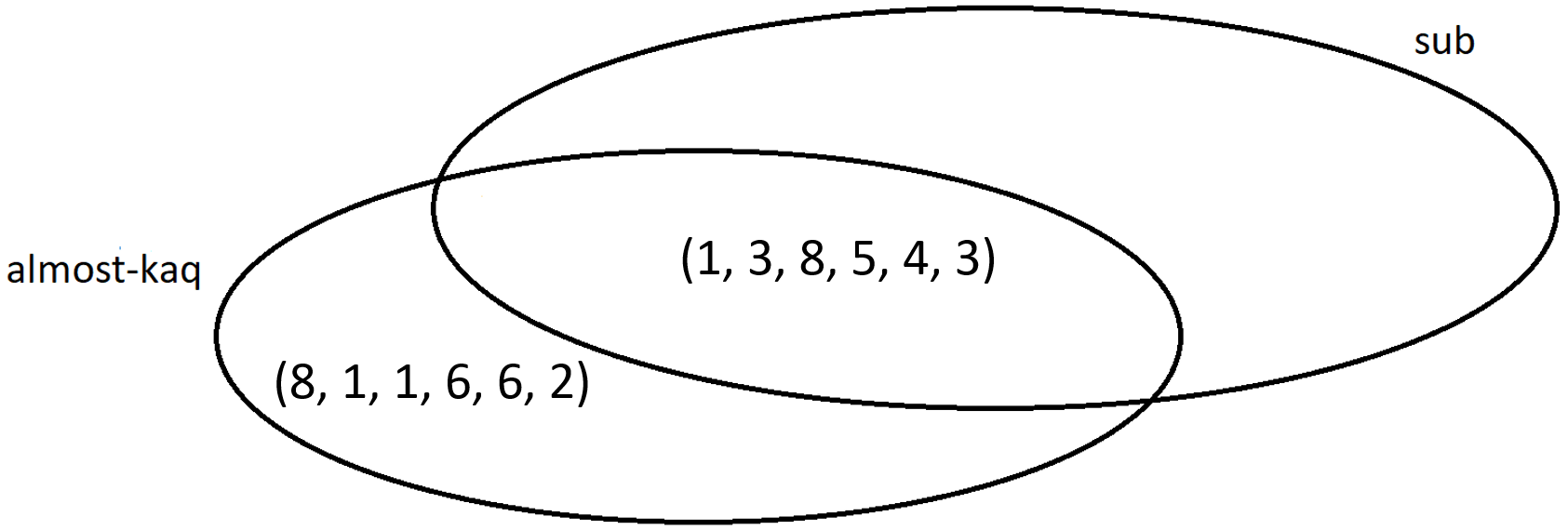}
    \caption{$\mathfrak{su}(5)$: Only $L_{26}$ results has almost-\textbf{kaq} patterns (Tables \ref{l24results5},\ref{l26results5}). $(8,1,1,6,2)$ is not \textbf{sub}, while $(1,3,8,5,4,3)$ is \textbf{sub} by the combination of the $1D+3D+3D$ eigenspaces.}
    \label{l246su5}
\end{figure}

\begin{figure}[h]
    \centering
    \includegraphics[trim=3cm 0 0 0, scale = 0.25]{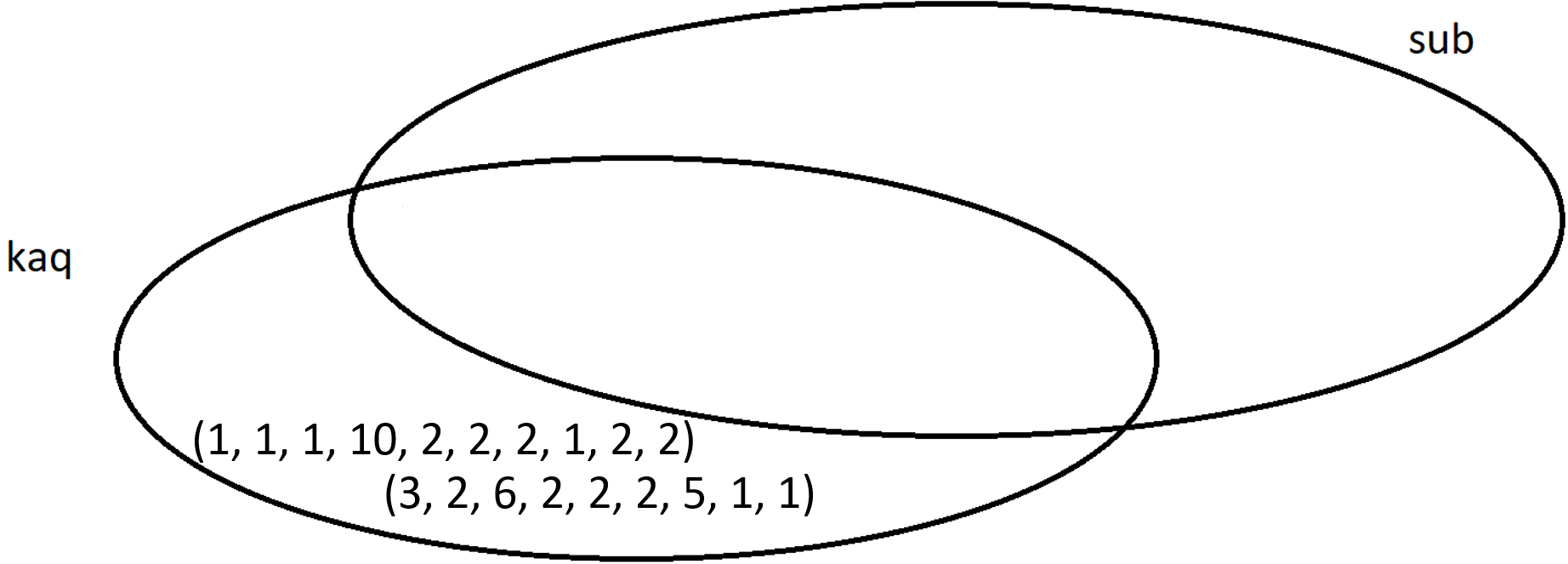}
    \includegraphics[trim=-6cm 0 0 0, scale = 0.25]{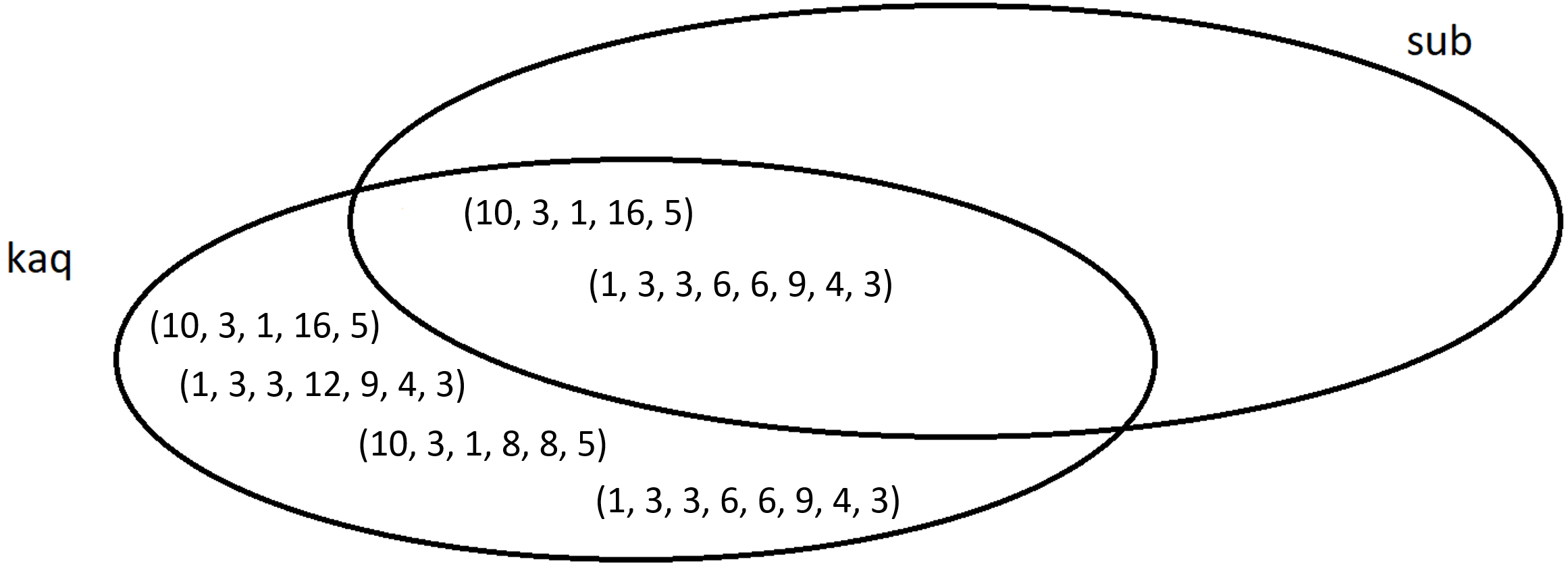}
    \caption{$\mathfrak{su}(6)$: There are more non-\textbf{sub} patterns than \textbf{sub}. Using \textbf{tensor} basis, we obtain only non-\textbf{sub} solutions such as $(10,3,1,16,5)$, however the same pattern obtained by using Gell-Mann basis, gives only \textbf{sub} solutions (by the combination of $10D+3D+1D$ or $10D+3D$ eigenspaces), hence why this pattern appears twice.}
    \label{l246su6}
\end{figure}

\begin{figure}[h]
    \centering
    \includegraphics[trim=3cm 0 0 0, scale = 0.25]{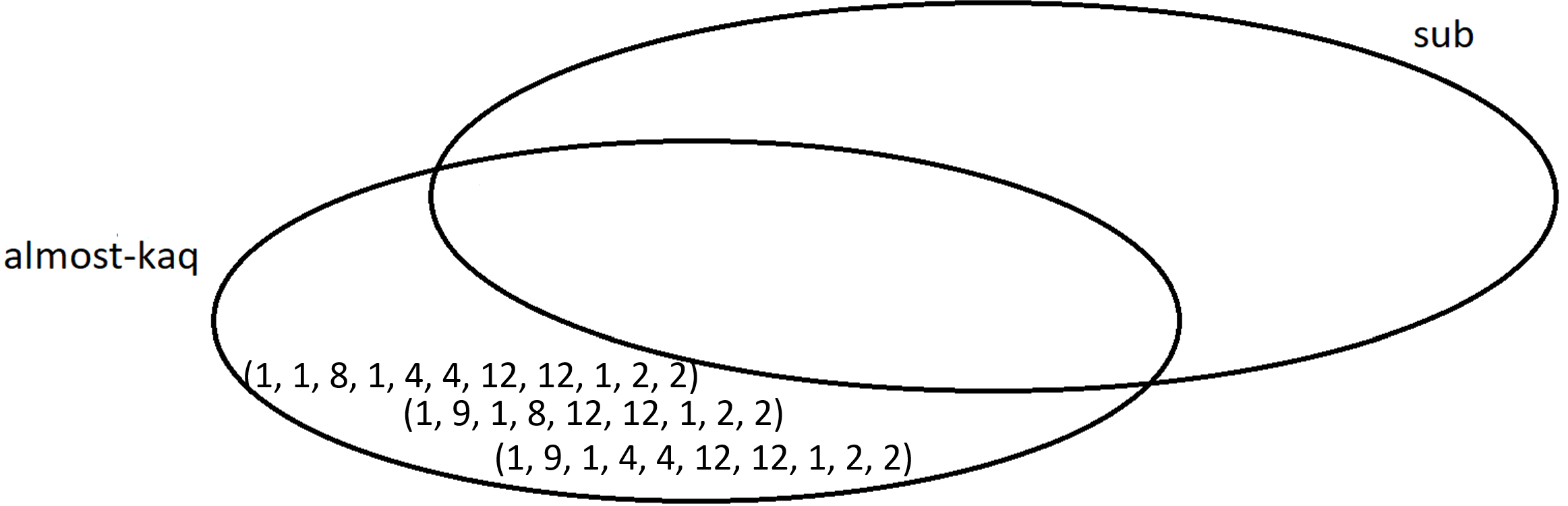}
    \includegraphics[trim=-5cm 0 0 0, scale = 0.25]{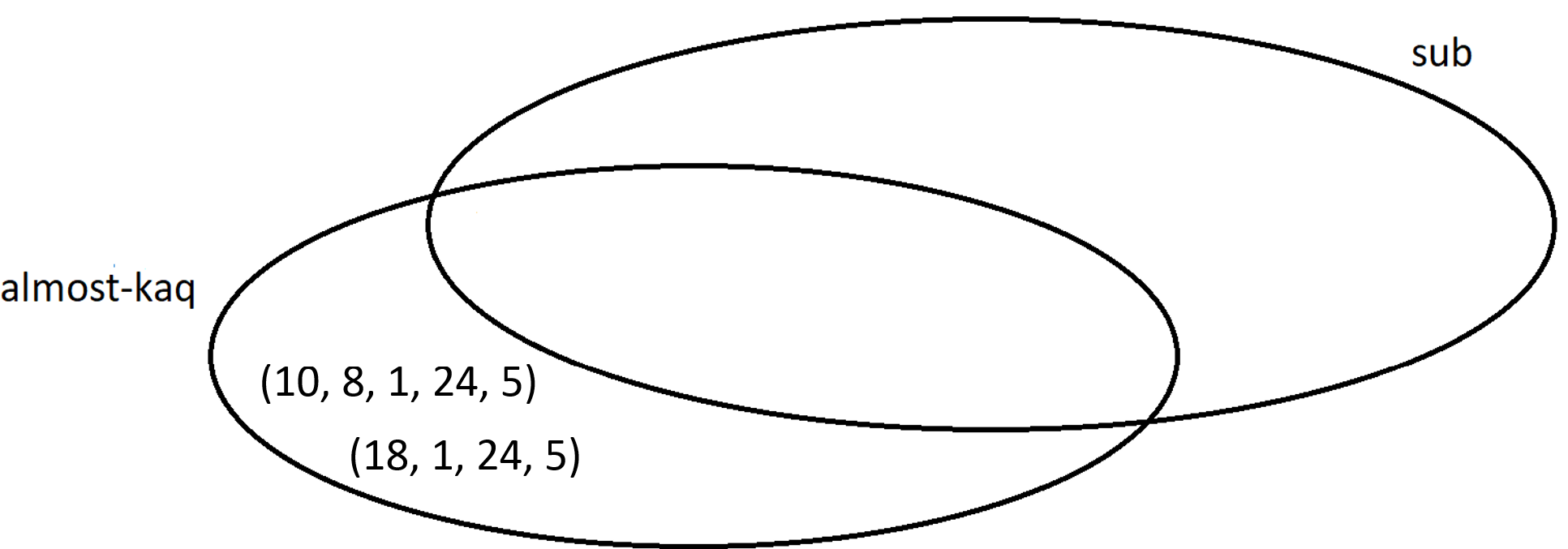}
    \caption{$\mathfrak{su}(7)$: All almost-\textbf{kaq} patterns were non-\textbf{sub}.}
    \label{l246su7}
\end{figure}

\begin{figure}[h]
    \centering
    \includegraphics[trim=5cm 0 0 0, scale = 0.25]{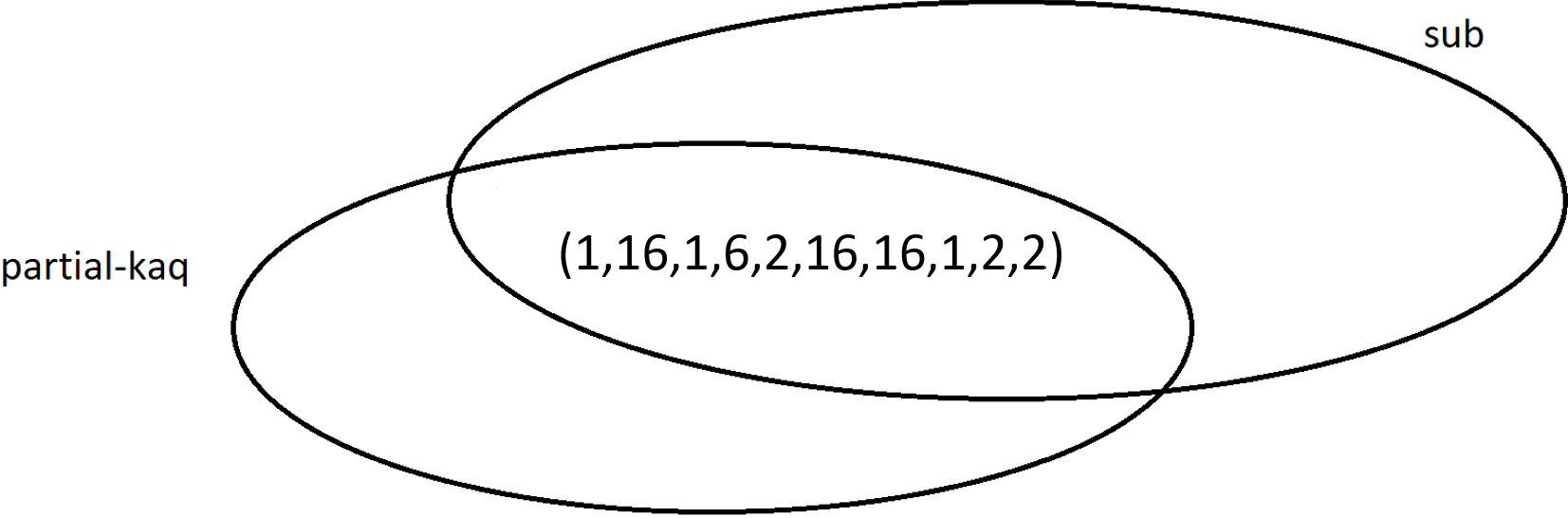}
    \includegraphics[trim=-5cm 0 0 0, scale = 0.25]{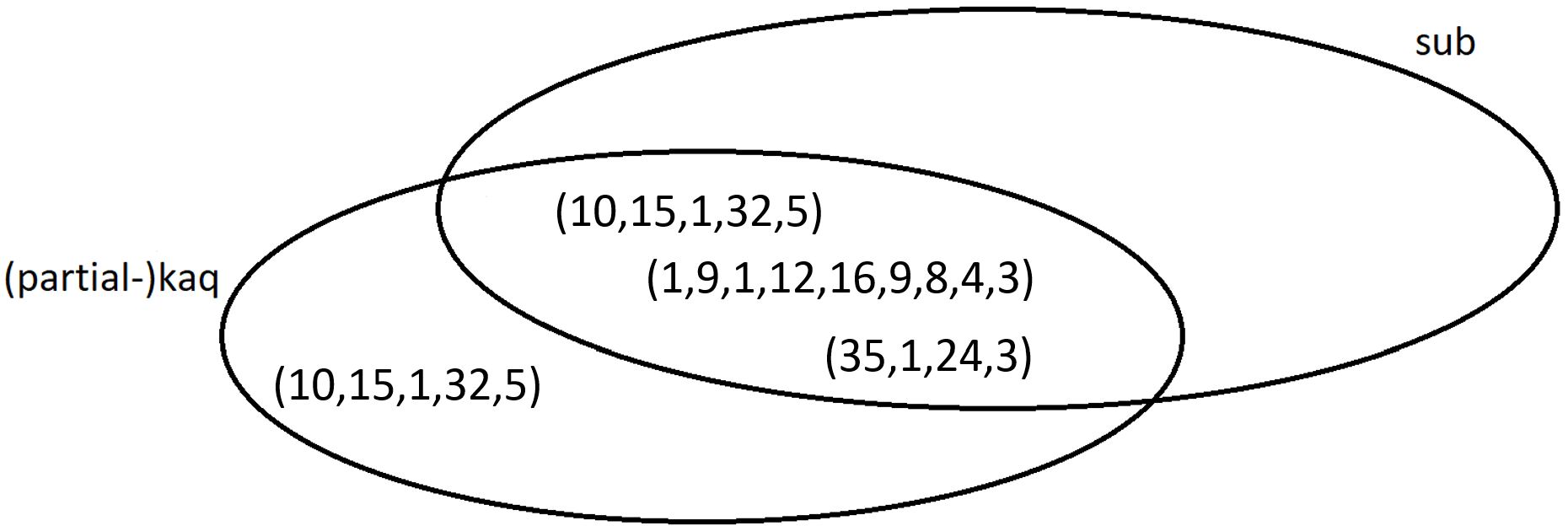}
    \caption{$\mathfrak{su}(8)$: On the left ($L_{24}$), we have one degeneracy pattern which is partial-\textbf{kaq} and all its instances are \textbf{sub}. On the right ($L_{26}$), the degeneracy pattern $(10,15,1,32,5)$ for $k=500$ has 11 \textbf{kaq} instances and 17 partial-\textbf{kaq} instances (which obviously includes the previous 11), \textit{all} of which are non-\textbf{sub}. However the same pattern for $k=1000$, with only partial-\textbf{kaq} results, is \textbf{sub} (by the combination of $10D+15D$ eigenspaces), along with all the other partial-\textbf{kaq} solutions for $k=1000$.}
    \label{l246su8}
\end{figure}

~
\\
\indent Some local minima are associated with Lie subalgebras. This was observed in \cite{freedman2021universe}, where we defined a solution to belong to \textbf{sub}$_n$ when a Lie \textbf{sub}algebra can be found that corresponds to one (or a combination of some) of its eigenspaces.  It is important to disambiguate generic \textbf{kaq} minima from those associated with Lie subalgebras as the latter may represent a distinct symmetry breaking process. Here, we show in Figures (\ref{l246su4},\ref{l246su5},\ref{l246su6},\ref{l246su7},\ref{l246su8}) which of our almost/partial-\textbf{kaq} solutions are also \textbf{sub}$_n$. We were previously \cite{freedman2021universe} able to find \textbf{kaq} solutions which are not \textbf{sub} using diagonal initialization of gradient flow (\textbf{DiagPerturbId}), however as the figures show, there are also many such examples for $n=4,6,8$ using the \textbf{GenPerturbId} method.

In each figure, solutions of $L_{24}$ (left) and $L_{26}$ (right) are put into a Veen diagram. Some solutions appear twice because they were instances of local minima with the same degeneracy pattern where one was \textbf{sub}$_n$ and the other was not. 

For each solution $g$, first, we computed its eigenbasis $\{iH_1,\ldots,iH_{n^2-1}\}$ with eigenspaces $\mathfrak{su}(n)=\oplus_{i=1}^t E_i$. Then, given a proper vector space $V = \oplus_{i \in \mcI} E_i$ with $\dim V >1$, formed by a subset $\mcI$ of $g$ eigenspaces, of which there are $2^t-2-|\{i|\dim E_i = 1\}|$, we computed the distance of the bracket of every two elements from the basis of $\oplus_{i \in \mcI} E_i$ to $V$ itself. If all distances were less than $1e-2$, $g$ was confirmed as an instance of \textbf{sub}. Otherwise, if no combination yielded a Lie subalgebra, $g$ was classified outside of the \textbf{sub} diagram. The same approach was taken in \cite[Figures 9-10]{freedman2021universe}.


\subsection{\textit{b}-score for \texorpdfstring{$\mathfrak{su(8)}$}{su(8)} solutions}\label{bscoresection}~
\\
\indent We compute the ``$b$-score'' or ``body-number-score'' of the \textbf{kaq} solutions found for $\mathfrak{su}(8)$ (see the motivation in \cref{bscoredfn1}). Notice that such solutions were only found in the $L_{26}$ results in \cref{l26results8}. There were $11$ many for $k=500$ and one for $k=1000$. The goal here is to look for the structure of the Brown-Susskind penalty metrics emerging. In these metrics the norm of a $g$-principal direction decays exponentially with its weight or body-number. We see, if anything, only a weak signal at $n=8$.

\begin{figure}[h]
    \centering
    \includegraphics[scale = 0.5]{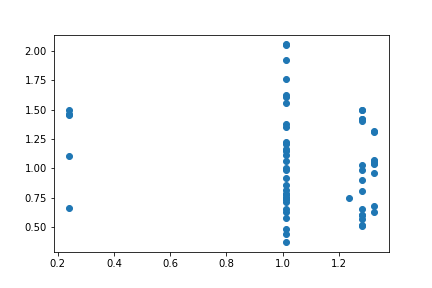}
    \caption{Pattern $(10,15,1,32,5)$ with $k=500$ and eigenvalues found by the gradient descent (before re-inverting $g$). $x$-axis is the eigenvalue which takes values at (1.32, 1.28, 1.236, 1.01, 0.24), and $y$-axis is the $b$-score. We see a modest negative ($-0.145$) correlation.}
    \label{bscorefig}
\end{figure}

The $b$-score for any $H_j = A_j \otimes B_j \otimes C_j$ decomposition is defined as $(1-\det(A_j))(1-\det(B_j))(1-\det(C_j))$  (note that $||A_j||_2=||B_j||_2=||C_j||_2=1$). We compute the $b$-score of every one of the $63$ eigenstate of $g$ for the $12$ solutions mentioned previously. We then compute the correlation of the $b$-score of $H_j$ with the eigenvalue of $g$ for $H_j$. The best result, correlation of $-0.145$, was for one of the 11 patterns $(10,15,1,32,5)$ with $k=500$. In \cref{bscorefig}, we show the scatter plot of the  $b$-score of $H_j$ and its eigenvalue. As the reader can see, the downward trend is only barely perceptible; a tendency toward Brown-Susskind geometrics is not yet confirmed. Indeed, other local minima metrics on su(8) show similarly weak trends but with the opposite sign.

\section{Summary and Outlook}
Through a numerical study of SSB from the GUE on $\mathfrak{su}(n)$, $n=4,5,6,7$, and 8, \textbf{kaq} and almost-\textbf{kaq} metrics (and their associated probability distributions), we find across two classes of natural functionals a convincing pattern. \textbf{Kaq} local minima are common for $n$ composite and almost-\textbf{kaq} local minima are common for $n$ prime. It appears that nature ``likes'' to organize large Hilbert spaces into tensor products of smaller ones, leaving a few (``rep\^echage'') dimensions to the side as necessary. This finding opens the door to number theory---being chiefly the study of primes---to enter the foundations of physics in a new way. In string theory \cite{smit87}, arithmetic structures on Riemannian surfaces provide a long-standing connection to number theory. The ``prime factor'' and ``leaky universe'' scenarios, described in \cref{introduction}, can now be added to this.

More generally, it should be incumbent on any foundational discussion such as ours to attempt contact with string theory. One way, as mentioned, could be through a common number theoretic context. Another is through geometry. We thank Greg Moore for the observation that, at least for the bosonic version, the Leech lattice could be the key to picking out the microscopic dimension of space-time. It is a natural goal, once interacting dof have appeared (as they now have) to see if they naturally organize themselves into a lattice geometry, perhaps even Leech-like (as opposed to e.g.\ a complete graph). This is utterly beyond naive quantum simulation, but could perhaps be approached with the help of an effective model, the analogy of Crick, Watson and Franklin studying DNA with a ball and stick model.

With more powerful computational resources and better techniques (\cref{simulationhardness}), it might be possible to study SSB at $n=16$ to see if we can confirm the ``hint'' of penalty metric structure that we discussed at $n=8$ in the anti-correlation of the $b$-function with norm of the principal axes (\cref{bscoresection}).

We thank Adam Brown for a suggestion we hope to follow. Rather than looking only for the ``initial Hamiltonian,'' he suggests one should look for a triple: $(H_0, \psi_0, \operatorname{entropy}(t)) =$ (Hamiltonian, initial state, and the behavior of entropy growth on subsystems). There appears to be something magical in how the universe's $H_0$ and $\psi_0$ conspire to allow subsystem entropies to gradually and uniformly increase over billions of years. Our universe is decidedly not a ``Boltzmann brain,'' look under any rock and you will see entropy on the increase. It seems that we can adapt the discussion of functionals $f: \{\text{metrics on such } g_{ij}\} \ra \R$ to $\tld{f}: \{g_{ij}\} \times \{\text{initial }\psi_0\} \ra \R$ by treating $\psi_0$ as a \textbf{source} in the Feynman diagram. Then a local minimum $(g_{ij}, \psi_0)$ gives rise to a probability distribution $::e^{-g_{ij}H_0^iH_0^j}$ from which we draw $H_0$ to obtain the pair $(H_0, \psi_0)$. At least in the case of $\operatorname{su}(4)$ it appears quite realistic to study entropy growth $S(t)$ w.r.t.\ any \textbf{kaq} decomposition, $4 = 2 \times 2$. We would look for interesting transients in the behavior of $S(t)$, which might be manifest before its quasi-periodic nature dominates.
\section*{Acknowledgments}
The first named author would like to thank the Aspen Center for Physics for their hospitality. The second named author would like to acknowledge the support of the Perimeter Institute for Theoretical Physics and Microsoft. Research at Perimeter Institute is supported by the Government of Canada through Innovation, Science and Economic Development Canada and by the Province of Ontario through the Ministry of Research, Innovation and Science. The experiments were conducted using Microsoft computational resources.
\bibliographystyle{apa}
\bibliography{main}

\bigskip
\address{\textsuperscript{*\label{1}}
	Microsoft Research, Station Q, and Department of Mathematics, University of California, Santa Barbara, CA 93106, USA 
}

\address{\textsuperscript{$\dagger$\label{2}}
	Perimeter Institute for Theoretical Physics, Waterloo, ON N2L 2Y5, Canada \\
	\indent \textsuperscript{$\bullet$\label{3}} Research Consultant, Microsoft
}
\end{document}